# A Wideband Reconfigurable Intelligent Surface for 5G Millimeter-Wave Applications

Ruiqi Wang, Yiming Yang, *Student Member, IEEE,* Behrooz Makki, *Senior Member, IEEE,* and Atif Shamim, *Senior Member, IEEE*

*Abstract*—Despite the growing interest in reconfigurable intelligent surfaces (RISs) for millimeter-wave (mm-wave) bands, and the considerable theoretical work reported by the communication community, there is a limited number of published works demonstrating practical implementations and experimental results. To the authors' knowledge, no published literature has reported experimental results for RISs covering the n257 and n258 mm-wave bands. In this work, we propose a novel wideband RIS design that covers the entire mm-wave 5G n257 and n258 bands. In simulations, the unit cell can maintain a phase difference of 180° ± 20° and a reflection magnitude greater than -2.8 dB within 22.7 to 30.5 GHz (29.3% bandwidth) using one-bit PIN switches. The proposed unit cell design with four circular cutouts and long vias could realize wideband performance by exciting two adjacent high-order resonances ($2.5f$ and $3.5f$). The periodic unit cells can maintain an angular stability of ± 30°. Based on the proposed unit cell, a 20×20 RIS array is designed and fabricated with a size of $7.1\lambda \times 7.1\lambda$. The measurement results demonstrate that the proposed RIS could maintain a 3 dB peak gain variation bandwidth among various array configurations within 22.5 to 29.5 GHz (26.9%) and with a beam scanning capability of 50°, making this design a good candidate for 5G mm-wave applications.

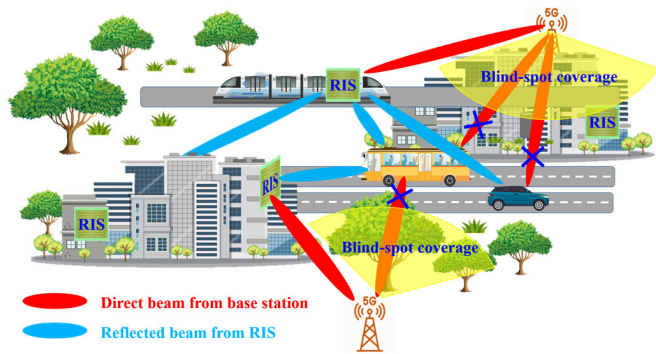

Fig. 1. Conceptual scheme of the reconfigurable intelligent surface (RIS) in practical applications.

## I. INTRODUCTION

RECENLTY, fifth-generation (5G) millimeter-wave (mm-wave) communication systems have received significant commercial and research attention due to their sizeable available bandwidth [1-5]. Although the 5G mm-wave system could significantly improve the data rate and communication capacity, the electromagnetic (EM) wave propagation suffers from a higher atmospheric attenuation. Thus, at such frequencies, the EM waves can get severely affected or even blocked by obstacles, causing signal blind spots and restricting the system to line of sight (LOS) communication. Such a problem is even more visible beyond the 5G network with increased quality-of-service requirements and higher frequencies. A reconfigurable intelligent surface (RIS) is beneficial to solve this problem because it can help establish a secondary LOS to maintain the communication link when a blockage occurs in the direct path between the transmitter (Tx) and receiver (Rx) [6]. A conceptual scheme of the RIS in a practical urban environment with multiple signal blind spots is depicted in Fig. 1, where the RIS is able to provide a second LOS to address the blind spots. This RIS-based wireless communication is controllable and programmable, which means that there can be unprecedented opportunities for improving the quality of service for wireless communication systems [6].

Despite the great performance enhancement promise of RISs, there is not much experimental evidence in published literature supports these claims. Most published works focus on the ideal RIS and are limited to theoretical analyses. For example, [7] reported an ideal and theoretical RIS for energy-efficient downlink multiuser communication applications. Similarly, [8] proposed the concept of RIS-assisted communications to the realm of index modulation. In addition, [9] and [10] reported on optimization methods for RIS beamforming for wireless network designs.

However, published literature related to practical RIS designs and their experimental characterization, particularly for mm-wave bands, is scarce. Out of the experimental studies, majority have been conducted on the sub-6 GHz band [11–14], and some operate at around 10 GHz [15], [16]. Few reports have addressed practical RIS designs operating at mm-wave bands. For example, [17] proposed a practical RIS design and performed near- and far-field measurements for signal enhancement. However, the working bandwidth for the design

This work is supported by Ericsson Research under Grant OSR#4606.
Ruiqi Wang, Yiming Yang, and Atif Shamim are with Computer, Electrical and Mathematical Sciences and Engineering Division, King Abdullah University of Science and Technology (KAUST), Thuwal, 23955-6900, Saudi Arabia (email: ruiqi.wang.1@kaust.edu.sa, yiming.yang@kaust.edu.sa, atif.shamim@kaust.edu.sa).
Behrooz Makki is with Ericsson Research, Ericsson, 417 56 Gothenburg, Sweden (email: behrooz.makki@ericsson.com)

[17] is from 27.5 to 29.5 GHz (7% bandwidth), which cannot fulfill the large bandwidth requirements for 5G mm-wave band. Shekhawat et al. [18], [19] designed a compact unit cell and a 25×32-element RIS prototype working at 28.5 GHz, but the operation bandwidth was not provided. Rains et al. [20] proposed a unit cell design based on varactors. However, the results were obtained from a simulation, and no experimental results were demonstrated. Moreover, the bandwidth is still narrow, so the unit cell phase shift within the 180° ± 20° can be maintained between 24.7 and 26.6 GHz (7.5%). The 5G mm-wave frequency bands contain n257 (26.5 to 29.5 GHz), n258 (24.25 to 27.5 GHz), n259 (39.5 to 43.5 GHz), n260 (37 to 40 GHz) and n261 (27.5 to 28.35 GHz) bands. Therefore, the designed RIS for the n257 and n258 bands (n261 is within the n257 band) needs to cover the bandwidth from 24.25 to 29.5 GHz (19.5%), which is challenging and there is no previously reported work that has realized wideband PIN-switch-based RIS to cover this 5G mm-wave band (24.25 to 29.5 GHz).

In this work, we designed, fabricated, and tested a novel 5G mm-wave RIS. A novel unit cell design of patches with four circular cutouts is proposed to satisfy the sizeable bandwidth requirements for the 5G mm-wave band. The proposed unit cell demonstrates wideband performance by exciting two adjacent high-order resonant modes ($2.5f$ and $3.5f$) with the help of four circular cutouts on the edge of the patch element, fully covering the 5G mm-wave n257 and n258 bands (24.25 to 29.5 GHz). The design maintains a simple structure with three layers (patch, ground, and biasing network layers). The PIN switches are designed on the backside of the RIS, avoiding interference from diodes and soldering pads. The designed unit cell structure provides adequate space for biasing line routing. A final 20×20 array with a size of 7.1λ×7.1λ is characterized using a two-horn measurement setup. The result reveal a 3 dB gain bandwidth from 22.5 to 29.5 GHz (26.9%) and a 50° beam-scanning range.

## II. Unit Cell Design and Simulation

Generally, the RIS is constructed using an array of unit cell elements with tunable electromagnetic resonators, which could modify the reflective waves by controlling the characteristic of each unit cell, such as the resonant frequency, reflection amplitude, phase, and polarization. Meanwhile, the RIS design should be distinguished from reconfigurable reflectarray designs [21-24] because reflectarrays have a fixed incident feed location and generally work in the near-field region for the feeding source antenna. Thus, reflectarray designs optimize the feed antenna and reflectarray elements to achieve optimum performance. The RIS, however, mainly concentrates on the reconfigurable element design and the surface distribution.

In this work, the structure of the proposed unit cell design is depicted in Fig. 2. The top metallic layer is a square patch whose side length is 3 mm. There are four 1.1 mm radius circular cutouts at its four edges, and the total length is 3.85 mm. Two side edges of the patch element are connected with vias (0.1 mm radius), which pass through two circular cutouts (0.35 mm radius) on the middle ground to attach the feeding network at the bottom layer. The top substrate is a 1.575 mm Rogers 5880 ($\varepsilon_r$ = 2.2, tan $\delta$ = 0.0009) which is low loss and suitable for 5G mm-wave application. Behind the middle ground plane layer is the bottom substrate with a 0.2 mm FR4 ($\varepsilon_r$ = 4.4, tan $\delta$ = 0.02) which has less effect on unit cell performance. Two substrate layers are laminated by a 0.1 mm FR4 prepreg sheet ($\varepsilon_r$ = 4.4, tan $\delta$ = 0.02). The PIN switches are the commercially available type MADP-000907-14020P in MACOM, which displays decent performance in the mm-wave band. One side of the PIN switch is connected with the middle ground through a short via, whereas the other side of the switch is connected with the top patch layer by a long via. By adjusting the biasing current, the PIN switch could work in two different states, ON and OFF, to achieve reconfigurability. A radio frequency (RF) choke is employed at the end of the long via, ensuring that the unit cell RF performance is not affected by the direct current (DC) feeding lines.

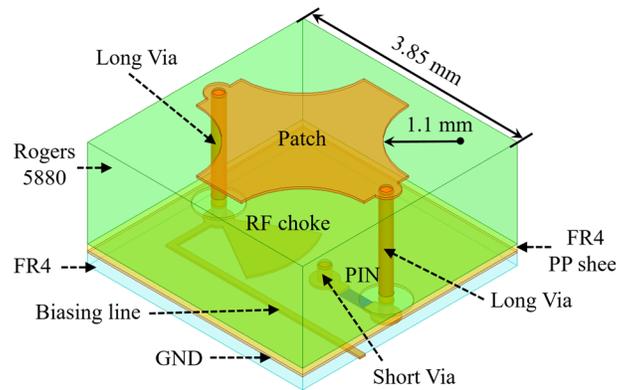

Fig. 2. Proposed unit cell design.

Some innovations of the proposed unit cell design are as follows. First, the proposed unit cell design possesses wideband performance due to the incorporation of circular cutouts into the square patch compared with the traditional square patch. The $S_{11}$ phase performance of the proposed unit cell design and the typical square patch unit cell when the PIN switches are in the ON/OFF states is illustrated in Fig. 3. It can be observed that the proposed unit cell with circular cutouts demonstrates steady slope for both of the ON and OFF states (Fig. 3(a)), whereas the traditional square patch unit cell has a nonuniform phase variation for ON/OFF states (Fig. 3(b)). Specifically, the 180°±20° phase difference between the ON and OFF states of the proposed unit cell design could be maintained in a wide band from 22.7 to 30.5 GHz (29.3%) and the reflection magnitude is greater than -2.8 dB in this bandwidth, as demonstrated in Fig. 4. The wideband characteristic is attributed to two adjacent resonant modes of $2.5f$ (23.1 GHz) and $3.5f$ (31.1 GHz) are excited in the ON state.

The detailed surface current distribution of the proposed cell structure is depicted in Fig. 5, where the surface current on the patch element is primarily distributed along the edges of circular cutouts. The dimensions of the circular cuts play critical roles in facilitating the excitation of these two modes. The proposed configuration can be compared with the simulated patch unit cell in Fig. 6, where only one resonant $3.5f$ mode occurs in the ON state without the circular cutouts. Additionally, the RF chokes and PIN switches are designed under the patch, saving the space near the unit cell boundaries for the DC biasing lines. Therefore, all biasing lines can be routed on the same layer as the PIN switches without the needs of adding other layers for multiple biasing lines. The proposed unit cell has wideband performance generated by exciting two resonant modes through a single layer with a simpler structure. Last, the PIN diodes are placed at the backside of the middle ground plane to avoid interference with the patch reflection. The circular cuts and the PIN placement achieves a compact unit cell size of 0.34λ×0.34λ (3.85×3.85 mm) compared with the standard half-wavelength unit cell design. The compact unit cell design could accommodate more elements in the same physical size, which is beneficial to practical miniaturization requirement.

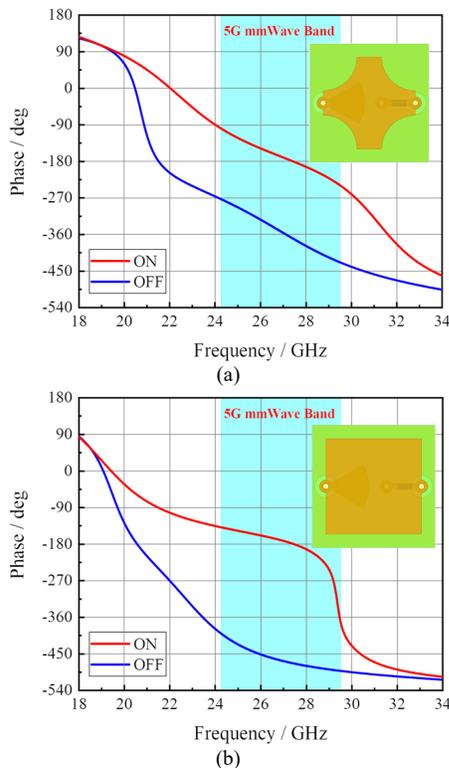

Fig. 3. The S₁₁ phase of the unit cell element when the PIN switch in the ON/OFF states. (a) Proposed unit cell. (b) Traditional square patch unit cell.

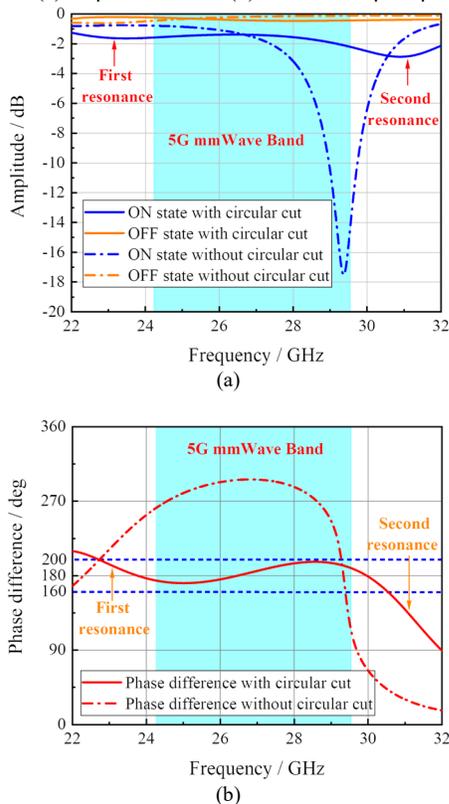

Fig. 4. Performance comparison of the patch unit cell with and without circular cuts. (a) Reflection magnitude. (b) Phase difference.

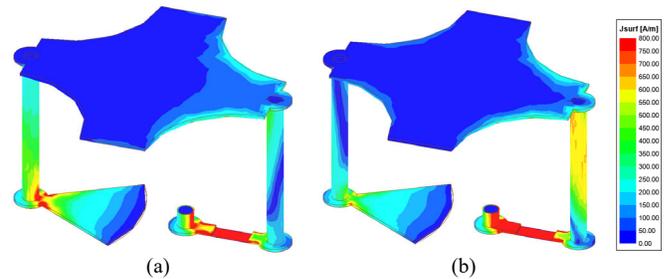

Fig. 5. Surface current distribution of the proposed unit cell in the ON state. (a) 2.5$f$ (23.1 GHz). (b) 3.5$f$ (31.1 GHz).

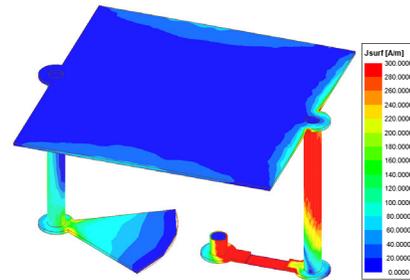

Fig. 6. Surface current of the patch unit cell in the ON state at 3.5$f$ mode (29.4 GHz).

In practical applications, the incident EM wave direction is arbitrary, requiring the RIS design should have angular stability to deal with incident waves from different directions. The reflection magnitude and phase difference of the proposed design for different incident angles are illustrated in Fig. 7. In the desired 5G mm-wave band (24.25 to 29.5 GHz), the unit cell has a reflection magnitude greater than -2.5 dB, and the phase difference can be maintained from 140° to 200° when incident wave direction varies between 0° and 30°.

## III. RIS Design and Simulation

The proposed RIS array is demonstrated in Fig. 8, comprising 400 unit cells (20×20). As mentioned above, the designed unit cell structure provides a sufficient space for the biasing lines to extend to the board edges. Twenty connectors (DF40TC(3.0)-20DS) were soldered onto the two edges of the printed circuit board (PCB) to connect the biasing lines to the control circuit board behind it. The total array size is 77×77 mm, corresponding to 7.1λ×7.1λ, where λ is the free-space wavelength at the frequency of 27.5 GHz. With additional connectors on the edge of the array, the PCB size is 96×96 mm.

A detailed full-wave simulation model in High-Frequency Simulation Software (HFSS) is depicted in Fig. 9. Each unit element of the proposed RIS is configured based on the discretized phase distribution calculated through the following

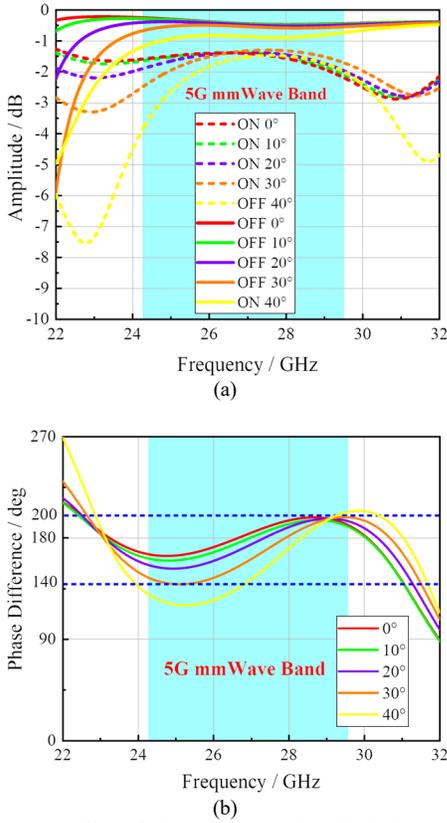

Fig. 7. Angular stability of the proposed unit cell design. (a) Reflection magnitude. (b) Phase difference.

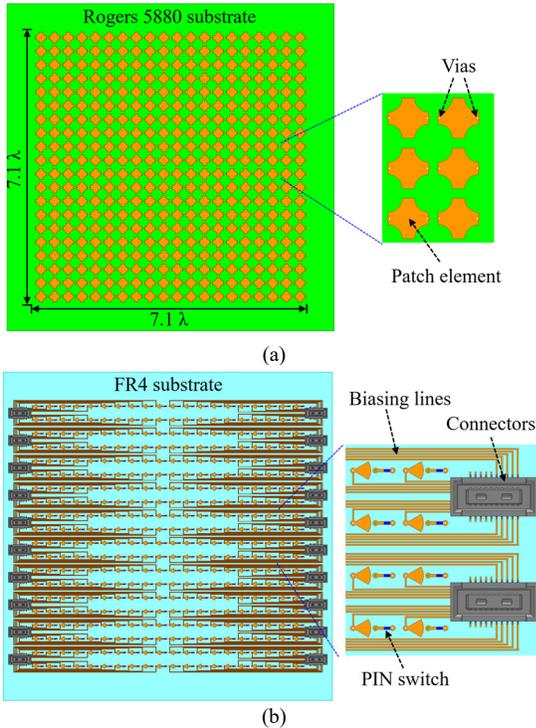

Fig. 8. Array configuration of the proposed RIS. (a) Front view. (b) Back view.

equation [17]:

$$\varphi_{ij} = k|\vec{r}_{ij}^{\,e} - \vec{r}^{\,f}| - k \cdot (\vec{u}_0 \cdot \vec{r}_{ij}^{\,e}) + \Delta\varphi \quad (1)$$

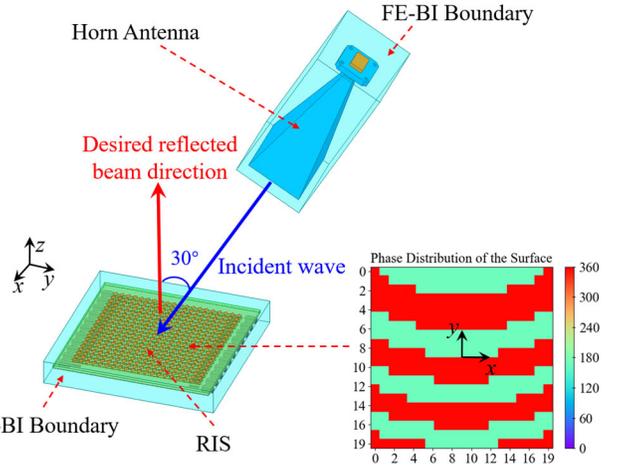

Fig. 9. Full-wave simulation of the proposed RIS.

where $\varphi_{ij}$ is the ideal phase value for the unit cell at $(i, j)$, $i,j =$ 1...20 in this case, $k$ denotes the wavenumber at the free space, $r_{ij}^{e}$ is the element position vector, and $r^{f}$ represents the feed position vector, $u_0$ indicates the unit vector of the desired reflected beam direction, and $\Delta\varphi$ denotes a constant reference phase. The proposed design has only two states by switching the ON and OFF status of the PIN diodes; therefore, the continuous ideal phase distribution should be quantized as two states using the following criterion:

$$\varphi_{ij} = \begin{cases} \varphi_{on}, |\varphi_{ij} - \varphi_{on}| \leq |\varphi_{ij} - \varphi_{off}| \\ \varphi_{off}, |\varphi_{ij} - \varphi_{on}| > |\varphi_{ij} - \varphi_{off}| \end{cases} \quad (2)$$

where $\varphi_{on}$ is the unit cell phase value when the PIN switch is ON, and $\varphi_{off}$ is the phase value when the PIN is OFF. These two values can be obtained from the HFSS simulated results of the proposed unit cell. It should be noted that $\varphi_{on}$ and $\varphi_{off}$ have different values at the entire frequency band, but the difference between them is 180°±20°.

According to the simulated angular stability of the proposed RIS, it has decent performance with the incident angle varying from -30° to +30°. Therefore, in Fig. 9, the incident wave direction is set as -30°, transmitted from a WR-34 standard 20 dBi gain horn antenna (PE9851B/SF-20) from 22 to 33 GHz. The phase distribution of the RIS is configured using (1) and (2), with $u_0$ (0,0,1). In the simulation model, the RIS and standard horn are settled in two boxes with finite element boundary integral (FE-BI) to reduce the computational cost of the free space between the horn and the RIS. The simulated three-dimensional (3D) radiation pattern of the proposed RIS at 27.5 GHz is presented in Fig. 10, with a peak gain of 20.9 dBi at 0° reflection, validating the reconfigurability of the designed RIS.

The switch control circuit schematics, control circuit PCB, and flexible printed circuit (FPC) connectors are illustrated in Fig. 11. The RIS array board and control circuits are connected though the FPC. The overall RIS is driven by the ESP-WROOM-32 module, which runs a serial communication port for the user interface and computes the array patterns. Attached to the controller are 50 8-bit phase registers (74HC595D), which can multiplex the controller IO and drive the PMOS in order to bias the 400 PIN switches. Each PIN diode of a unit cell is in series with an LED as visualization of the PIN working state for ease of debugging.

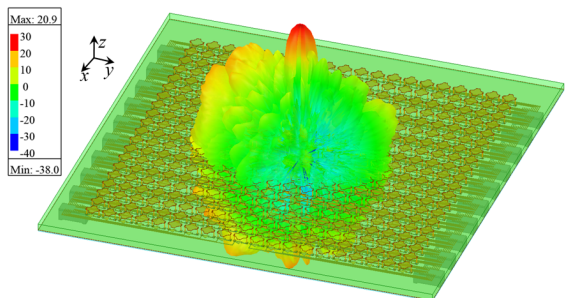

Fig. 10. The 3D radiation pattern of the RIS at 27.5 GHz.

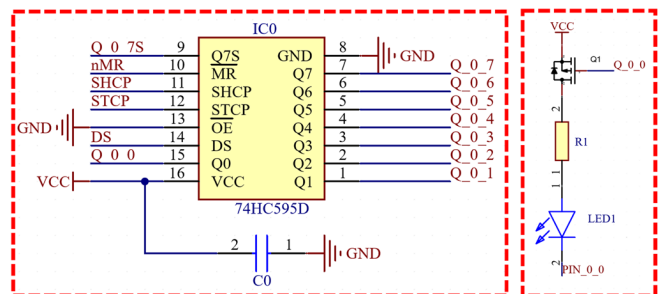

(a)

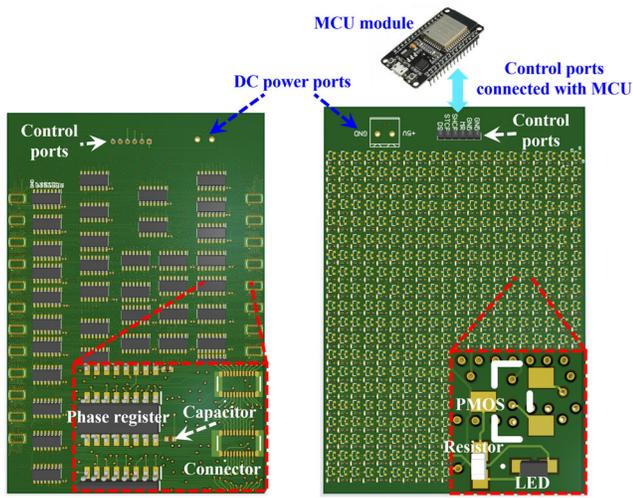

(b)

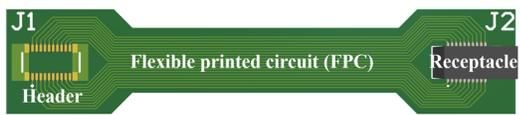

(c)

Fig. 11. Control circuit design. (a) Switch control and output schematic. (b) Front and back views of the control board. (c) Flexible printed circuit connectors.

## IV. Fabrication and Measurement

The fabricated prototypes of the proposed RIS array configuration, control circuits, and overall RIS design are exhibited in Fig. 12. The components of the designed RIS, including the PCB array, connectors, FPC, and control circuit board are packaged inside the mechanical supports, which are 3D-printed using a Raise3D Pro2 printer with polylactic acid filaments. The control circuits are placed at the back side of the package, where the LED arrays demonstrate the ON/OFF state of each unit element. The power supply port contains a positive voltage (+3 to 5 V) and a ground input, which can be connected to a DC power source. Four control ports containing a clock signal, memory reset, and input data are driven by the ESP-WROOM-32 module, which is connected to a computer through USB to receive commands.

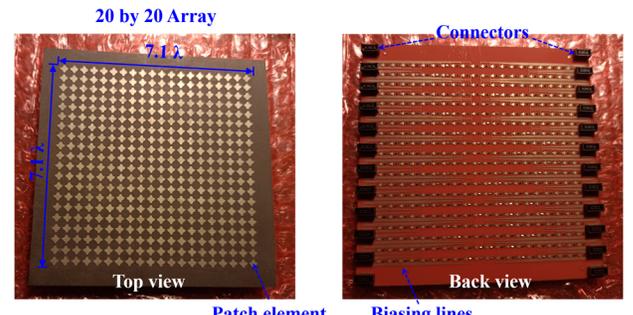

(a)

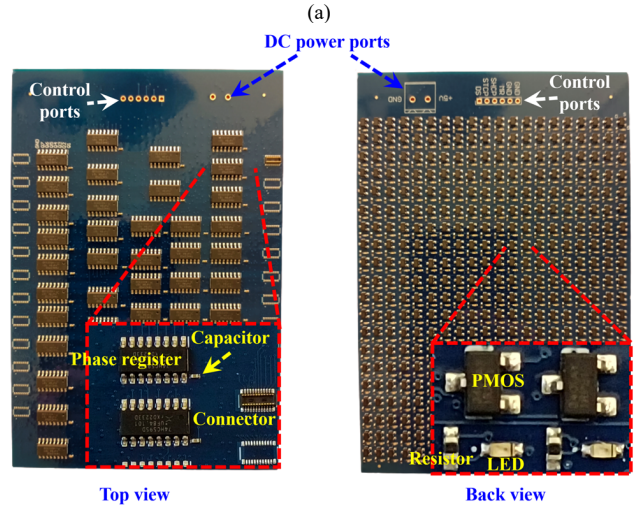

(b)

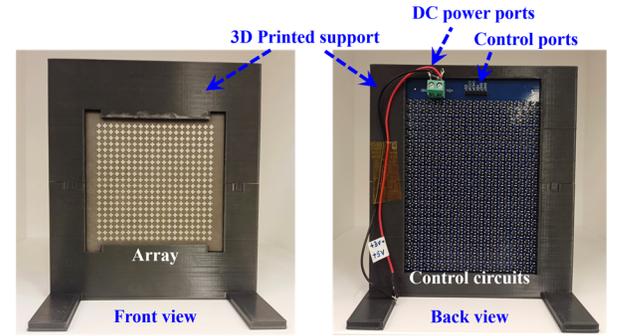

(c)

Fig. 12. Fabricated prototype of the proposed RIS design. (a) RIS array. (b) Control circuits. (c) RIS array packaged with control circuits board through FPC connectors.

The measurement setup of the proposed RIS is depicted in Fig. 13. Two horn antennas (PE9851B/SF-20) are utilized as the Tx and Rx in practical applications, which connect with an Anritsu ME7828A vector network analyzer (VNA). The RIS

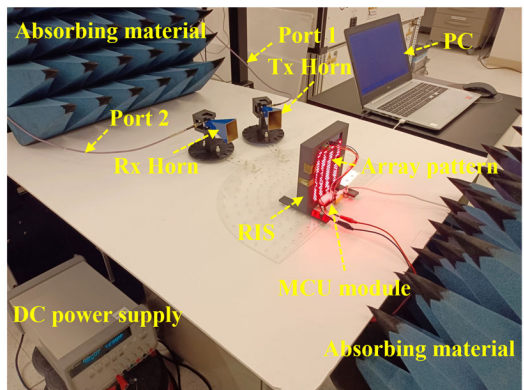

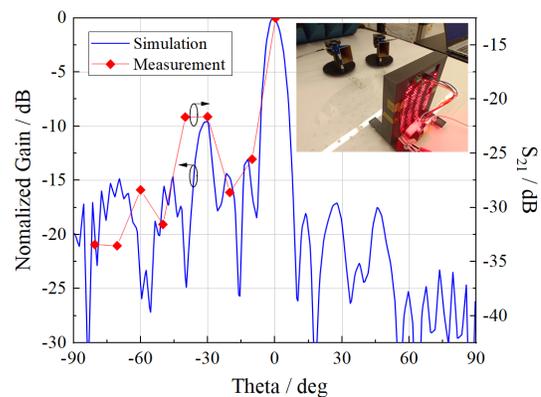

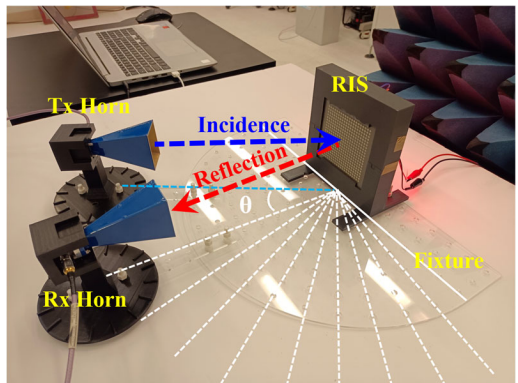

Fig. 13. Practical measurement setup of the proposed RIS design. (a)The overall measured system. (b) The detailed measured scheme between horns and RIS.

could reflect the incident beam to the configured beam direction instead of PEC reflection using a specific array pattern. In the measurement process, once the Tx horn position is determined, the RIS will generate the array pattern according to the previously discussed algorithm (1). It should be noted that the array has various patterns depending on the distance between the Tx horn and RIS operation frequency and the desired beam reflection direction. In this measurement, the Tx horn is placed at a distance of 20 cm from the RIS, which works in the near field of the horn antenna. To measure the reflection beam, we move the Rx horn from –80° to 0° with a 10° step while maintaining the same distance of 30 cm from the RIS. The measured $S_{21}$ parameter between the Tx and Rx horns represents the received signal power level related to the normalized gain of the RIS. Two pieces of absorbing material are placed at the front and back sides of the RIS measurement system to reduce the interference of the behind VNA and surrounding setup. The RIS could reflect the incident beam to the configured beam direction. When the array pattern is OFF, it has a similar reflected beam as a PEC reflection because all elements have a constant OFF reflection phase. The comparison between simulation and measurement is demonstrated in Fig. 14, which shows consistency that the measurement agrees well with the simulated results. The discrepancy between simulation and measurement is resulted from the printed support, control circuit, and fabrication and assembly errors of the RIS. It should be noted that the measured data are from –80° to 0° because the

Fig. 14. Comparison between simulated normalized gain and measured $S_{21}$ between Tx and Rx horns.

Rx horn overlaps with the Tx horn when the angle is greater than 0°.

Moreover, the comparison of the received signal strength level at 0° reflection (30° incidence) between turning ON and OFF the RIS pattern is demonstrated in Fig. 15. A significant improvement from –37.3 to –12.6 dB (24.7 dB enhancement) is observed at the center operation frequency of 27.5 GHz when the RIS pattern turns on compared with the OFF state. An average of 21.2 dB enhancement is obtained for the 5G mm-wave n257 and n258 frequency bands. It should be pointed out that each frequency point has its unique RIS pattern to achieve the optimum reflection performance. The RIS pattern in Fig. 15 is optimized for the 27.5 GHz frequency, where a peak gain is observed over the whole frequency range.

Detailed measurement procedures and results are discussed as follows. First, to validate the practical wideband performance of the proposed RIS design, the same measurements at different frequencies were conducted, as demonstrated in Fig. 16(a). Specifically, the incident wave and desired reflected wave directions were fixed. As discussed in the simulation section, the proposed RIS has angular stability with –30° to 30°. Thus, the Tx horn position is fixed, and the incidence angle is maintained at 30°, whereas the desired reflected beam direction is set to 0°. The measured setup remains the same, but working frequencies changed from 22.5 to 29.5 GHz with a 1 GHz step that each frequency point has a specific array pattern to reflect the 30° incident wave to 0°.

The measured signal strength of the desired beam direction (0° reflection) over a wide frequency range from 22.5 to 29.5 GHz is captured as demonstrated in Fig. 17. The measured $S_{21}$ parameter between the Tx and Rx horns changed from –12.5 to –14.7 dB within the whole measured frequency band. In other words, the proposed RIS design possesses a 3 dB peak gain bandwidth of over 26.9% which can fulfill the 5G mm-wave n257 and n258 bands requirements.

The detailed received signal level comparison of the RIS and PEC reflections is presented in Fig. 18(a–h). The array phase distribution (i.e., ON/OFF state of the PIN switches) at different operation frequencies is provided at the upper right corner of each figure. Uniform and consistent results were obtained within the whole frequency band from 22.5 to 29.5 GHz (26.9%), which proves its wideband reconfigurability. The measured side lobe level (SLL) values for the proposed RIS design are 7.9, 8.1, 4.7, 6.3, 7.1, 9.1, 8.4 and 8.7 dB for frequencies from 22.5 to 29.5 GHz. Therefore, an average of 7.5 dB for the SLL is obtained over the whole frequency range. The reason that the SLL at frequencies around 24.5 and 25.5 GHz is lower than at other frequencies is that the unit cell has a deteriorated phase difference (140°) at an incident angle of 30° (Fig. 7). Nevertheless, such a measurement proves that the proposed RIS design has wideband performance that could cover the 5G mm-wave n257 and n258 bands.

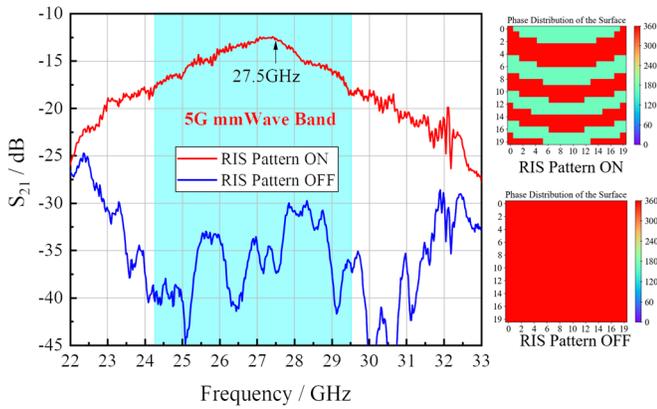

Fig. 15. Measured $S_{21}$ between Tx and Rx horns when turning ON/OFF the RIS pattern in the near field.

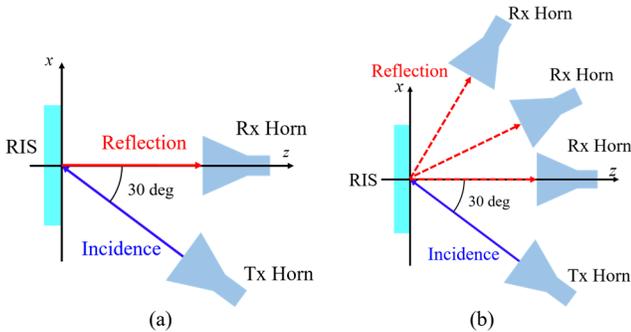

Fig. 16. Measurement scheme. (a) Wideband performance of the proposed RIS. (b) Beam scanning performance of the proposed RIS.

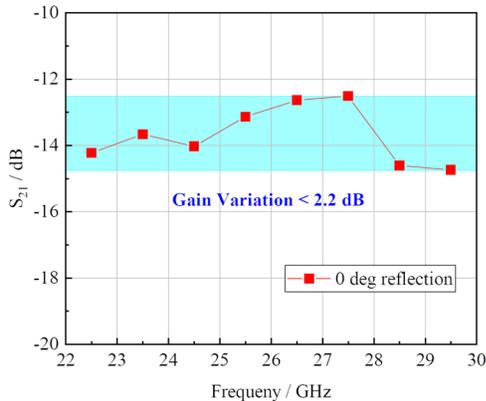

Fig. 17. Measured $S_{21}$ for the 0° reflection at different frequencies.

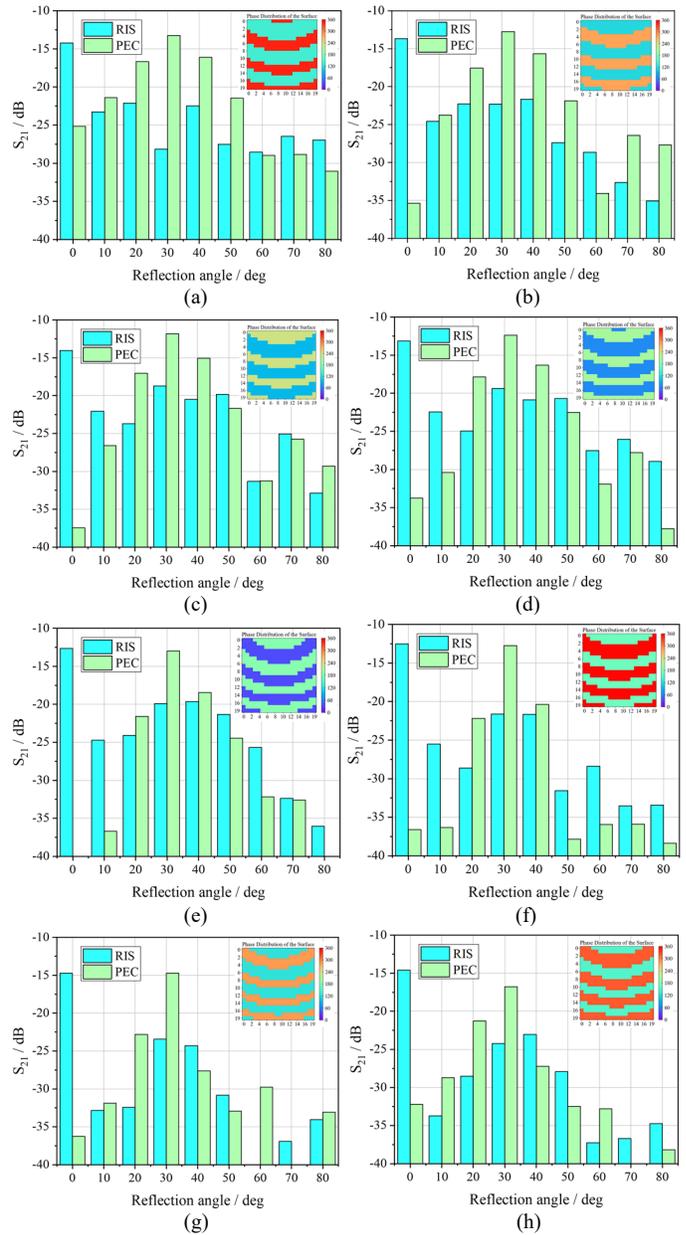

Fig. 18. Measured $S_{21}$ comparison of the RIS reflection and PEC reflection at different frequencies. (a) 22.5 GHz. (b) 23.5 GHz. (c) 24.5 GHz. (d) 25.5 GHz. (e) 26.5 GHz. (f) 27.5 GHz. (g) 28.5GHz. (h) 29.5 GHz.

Furthermore, another experiment on the beam scanning capability of the proposed design was also conducted. The detailed measured scheme is demonstrated in Fig. 16(b). In this case, the feed source location is the same as the previous experiment with the incident wave direction of 30°. In addition, the operation frequency is fixed at 27.5 GHz. However, this time, the desired beam reflection direction is changed from 0° to 50° to validate the beam scanning performance of the designed RIS. Therefore, the array pattern changes accordingly when with various reflected wave directions.

The measured $S_{21}$ values between the Tx and Rx horns at different reflected beam directions from 0° to 50° are presented in Fig. 19(a–f). The maximum reflected beam direction moves from the original 0° to 50° when varying the pattern configuration. The measured $S_{21}$ parameters between two horns for the desired beam reflection direction from 0° to 50° are –12.5, –12.8, –11.7, –11.3, –11.9, and –14.5 dB. Moreover, the SLL values for reflection angles from 0° to 50° are 9.1, 7.6, 11.3, 16.5, 12.9, and 5.8 dB. Thus, the proposed RIS design

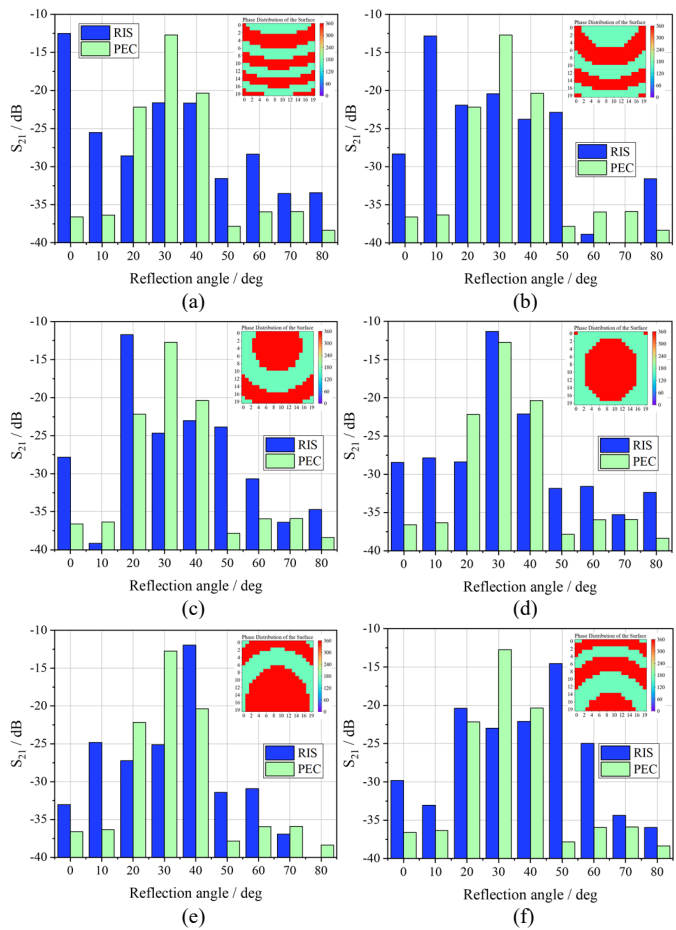

Fig. 19. Measured $S_{21}$ between the Tx and Rx horns at different reflected beam directions. (a) 0°. (b) 10°. (c) 20°. (d) 30°. (e) 40°. (f) 50°.

could demonstrate a beam steering capability of up to 50°.

The above-conducted measurements are the near-field tests, where the Tx horn is 20 cm away from the RIS. Therefore, the incident wave impinging on the RIS is a spherical wave. To validate the RIS performance in the far-field region, the Tx horn should be placed at a position that ensures the illumination on the RIS is a plane wave. In this experiment, the horn antenna (PE9851B/SF-20) has an aperture of 3.47×4.7 cm, so the far-field distance should be $2D^2/\lambda$ = 62.6 cm, where λ is the wavelength at 27.5 GHz. Hence, the Tx horn is moved to a distance of 70 cm to implement the far-field test, whereas the Rx horn is kept 30 cm from the RIS to maintain a relatively high received signal power level. The measurement setup has the same configuration as the near-field test in that the incident wave angle is 30°, and the desired reflection beam direction is 0°, which is achieved by controlling the RIS pattern (Fig. 16(a)).

The measured transmission between the Tx and Rx horns, when the RIS pattern is ON or OFF, is depicted in Fig. 20. The signal is enhanced substantially, and an average of 19.6 dB (from –42.6 to –23 dB) is obtained over the entire n257 and n258 bands. Although the signal enhancement has a 5.1 dB reduction compared with the near-field test, it is reasonable due to the higher path loss and less effective received area of the RIS compared with the horn placed in the near field. Nevertheless, the far-field test proved that the proposed RIS design could realize decent signal enhancement with a large bandwidth over the entire interested 5G bands. To compare the proposed work with previous practical RIS designs, the features of this work are summarized in Table I. Most RISs have a narrow bandwidth of less than 14% [11], [13], [14], [17]. As discussed, the previously published literature has primarily focused on the sub-6 band [11], [13], [14]. The work proposed in [17] aimed at a 5G mm-wave design, but the bandwidth is 7.3%, which cannot cover the entire 5G n257 and n258 bandwidths. In this work, the proposed design with a measured 26.9% bandwidth is the highest for the RIS operating at the 5G mm-wave band.

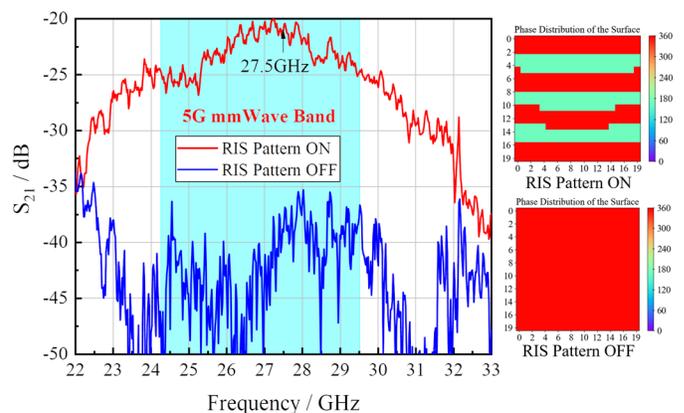

Fig. 20. Measured $S_{21}$ between Tx and Rx horns when turning ON/OFF the RIS pattern in the far field.

TABLE I PERFORMANCE COMPARISON WITH PREVIOUS RECONFIGURABLE INTELLIGENT SURFACE DESIGNS

| Ref. | Freq. (GHz) | Unit BW (%) | Size (λ²) | Gain BW (%) | Signal enhancement (dB) | |
|---|---|---|---|---|---|---|
| | | | | | Near field | Far field |
| [11] | 5.2 | 3.1 | 10.4× 10.4 | <5.8 | NA | ~15 (26 dB Directivity) |
| [13] | 3.15 | 14 | 3×3 | <14 | NA | NA |
| [14] | 5.8 | 8.7 | 15.5× 6.1 | <8.7 | NA | 27 |
| [17] | 28.5 | 7.0 | 9.5× 9.5 | 7.3 | ~30 | ~25 |
| This work | 27.5 | 29.3 | 7.1× 7.1 | 26.9 | 24.7 | 19.6 |

V. CONCLUSION

In this work, we designed a novel wideband RIS operating in the 5G mm-wave band, and the unit cell has a bandwidth of 29.3% (22.7 to 30.5 GHz). Wideband performance is realized by the two high-order resonances with 2.5$f$ and 3.5$f$ operation frequencies, which demonstrates a decent unit cell performance with incident angular stability of ±30°. A final RIS design with a 20×20 array configuration is fabricated and measured with a total size of 7.1λ×7.1λ revealing a 3 dB gain bandwidth of 26.9% (covering the whole 5G mm-wave n257 and n258 bands). The design can scan a reflection beam of up to 50°. Therefore, the proposed RIS design with wideband reflection characteristics is suitable for the 5G mm-wave communication system.